\documentclass[prd,twocolumn,notitlepage,superscriptaddress,nofootinbib,amssymb,aps,floatfix] {revtex4-1}
\usepackage{graphicx}
\usepackage{epstopdf}
\usepackage{amsmath}
\usepackage{amsfonts}
\usepackage{mathtools}
\usepackage{adjustbox}
\usepackage{amssymb}
\usepackage{dcolumn}
\usepackage{bm}
\usepackage{color}
\usepackage[linktoc=all]{hyperref}
\usepackage{cleveref}
\usepackage[section]{placeins}
\usepackage[freestanding,sicmds]{hepunits}
\usepackage{caption,subcaption}
\usepackage{ragged2e}
\numberwithin{equation}{section}

\newcommand{\ud}{\,\mathrm{d}}

\begin{document}
	
	\preprint{Imperial/TP/2021/MC/01}
	
	\title{Phase Transitions in de Sitter:\\
	Quantum Corrections}
	
	\author{Jos\'e Eliel Camargo-Molina} 
	\affiliation{Department of Physics and Astronomy, Uppsala University, Box 516, SE-751 20 Uppsala, Sweden}
	\author{Mariana Carrillo Gonz\'alez} 
	\affiliation{Theoretical Physics, Blackett Laboratory, Imperial College, London, SW7 2AZ, U.K}
	\author{Arttu Rajantie} 
	\affiliation{Theoretical Physics, Blackett Laboratory, Imperial College, London, SW7 2AZ, U.K}

	\date{\today}
	
	\begin{abstract}
	We investigate the decay rate of a false vacuum state in de Sitter space at high Hubble rates, using two methods: the Hawking-Moss instanton method which is fully quantum mechanical but relies on the saddle-point approximation, and the Starobinsky-Yokoyama stochastic approach which is non-perturbative but does not include quantum effects. We use the flux-over-population method to compute the Hawking-Moss decay rate at one-loop order, and demonstrate that in its domain of validity, it is reproduced by the stochastic calculation using the one-loop constraint effective potential. This suggests that the stochastic approach together with the constraint effective potential can be used to accurately describe vacuum decay beyond the saddle-point approximation.
	\end{abstract}
	
	\maketitle
	
	\section{Introduction}
	An ubiquitous phenomenon of quantum field theories is quantum tunneling which renders a classically stable vacuum metastable and leads to phase transitions. Precision calculations of the decay rate of such metastable vacuum are relevant for understanding and constraining possible physics beyond the standard model (SM) and non-minimal gravitational couplings. Given the current measurements of the SM Higgs and top quark masses \cite{CMS:2012qbp,ATLAS:2012yve}, our Universe seems to lie in a metastable state which should have a small enough decay rate \cite{Chigusa:2017dux,Chigusa:2018uuj,Rajantie:2016hkj,Espinosa:2020qtq,Branchina:2013jra}. In the early Universe the decay rate can be enhanced by different mechanisms \cite{Markkanen:2018pdo}; hence, rigorous calculations of such decay rates are important to understand the constraints on new physics.
	
	In flat space, a formal definition of the decay rate of the false vacuum is given by $\Gamma=-2\text{Im}(E)=2\text{Im}\left(\lim _{T \rightarrow \infty} (\ln Z) / T\right)$, where $Z$ is the path integral. The decay rate per unit volume can be computed using the saddle-point approximation which at next-to-leading (NLO) gives \cite{PhysRevD.15.2929,Callan:1977pt}
	\begin{align}
		\frac{\Gamma}{\mathcal{V}}=&\left(\frac{B}{2 \pi}\right)^{ 2}\left|\frac{\operatorname{det}^{\prime} S^{\prime \prime}\left(\phi_{b}\right)}{\operatorname{det} S^{\prime \prime}\left(\phi_{\mathrm{fv}}\right)}\right|^{-1 / 2} e^{-B_R}
		\label{eq:flatdecayrate}
	\end{align}
	where $\phi_b$ is the saddle-point or {\it bounce}, $\phi_{\mathrm{fv}}$ is the false vacuum, the prime on the determinant indicates that only non-zero modes are included, $B=S\left(\phi_{b}\right)-S\left(\phi_{\mathrm{fv}}\right)$, and $S$ is the Euclidean action. Note that the functional determinant ratio is divergent and can be regularized using standard QFT methods. A formal definition of the decay rate in curved spacetimes is not available, yet one can push forward by making analogies with the flat space result. Nevertheless, at high curvatures this analogy breaks down. In this letter we propose to take seriously the thermal interpretation of de Sitter (dS) to compute precise decay rates in the early Universe.
	
	\section{Decay rate in dS space}
	In the following, we will focus on decays from de Sitter to de Sitter spacetimes. The generalization of the decay rate computation to curved spacetimes was proposed in \cite{Coleman:1980aw}.  At small curvatures the Coleman-de Luccia bounce is expected to drive the decay, but as the curvature is increased a solution that has no flat space analog takes over, this is the Hawking-Moss instanton \cite{Hawking:1981fz}. At such large curvatures, the decay rate can also be computed through a different method dubbed the stochastic formalism \cite{Starobinsky:1986fx,Starobinsky:1994bd}. In the following we give a short review of both approaches. We will focus on computing the one-loop corrections through saddle-point approximation and comparing this result with the stochastic formalism one.
	
	\subsection{Stochastic approach}
	The stochastic formalism relies on splitting a light quantum field living in a dS space into long (classical) and short (quantum) modes, and describing the latter as stochastic noise. This framework has been shown to be useful for perturbative and non-perturbative quantum field theory computations in dS backgrounds, particularly in addressing issues for light fields \cite{Fumagalli:2019ohr,Moss:2016uix,Vennin:2015hra,Hardwick:2017fjo}. Assuming that the long-wavelength modes, $\phi$, satisfy an overdamped Langevin equation, it is found that the one-point probability distribution $P(t ; \phi)$ of $\phi$ at time $t$ follows the Fokker-Planck (FP) equation
	\begin{align}
		\frac{\partial \tilde{P}(t ; \phi)}{\partial t}&=\frac{3 H^{3}}{4 \pi^{2}} \tilde{D}_{\phi} \tilde{P}(t ; \phi) \ , \\
		\tilde{D}_{\phi}&=\frac{1}{2} \frac{\partial^{2}}{\partial \phi^{2}}-\frac{1}{2}\left(v^{\prime}(\phi)^{2}-v^{\prime \prime}(\phi)\right), \\
		v(\phi)&=\frac{4 \pi^{2}}{3 H^{4}} V(\phi) \ , \\
		\tilde{P}(t ; \phi)&=e^{\frac{4 \pi^{2} V(\phi)}{3 H^{4}}} P(t ; \phi) \ .
	\end{align}
	By expanding the probability in terms of the eigenfunctions of the FP equation, it can be found that the decay rate of the false vacuum is given by the lowest non-zero eigenvalue \cite{risken1989fpe,ElielandArttu}. 
	
	Note that this description is intrinsically classical. As we will show in this article, the thermally assisted tunneling approach suggests that one can use the constraint one-loop effective action in the stochastic approach to capture quantum corrections.
	\subsection{Thermally Assisted Tunneling}
	\subsubsection{Bounce solutions in de Sitter}
	When we include gravity and consider decays from de Sitter to de Sitter, the topology of the bounce solution is assumed to be a 4-sphere with the metric given by
	$ d s^{2}=d \xi^{2}+\rho(\xi)^{2} d \Omega_{3}^{2}$, where $\rho$ has zeros at $\xi=0$ and $\xi=\xi_{max}$. The bounce satisfies the boundary conditions $\phi'(0)=\phi'(\xi_{max})=0$, and while the field can approach the false vacuum, it actually never reaches it. For simplicity, we will consider a fixed de Sitter background, $\rho=\sin{\left(\xi H\right)}/ H$. This approximation is justified for small barriers where $\Delta V\equiv V\left(\phi_{\text {top }}\right)-V\left(\phi_{\mathrm{tv}}\right) \ll V\left(\phi_{\mathrm{tv}}\right)$. Here, we focus on the Hawking-Moss (HM) solution given by
	$\phi=\phi_\text{top}$. In analogy with flat space, the decay rate is given as
	\begin{equation}
		\frac{\Gamma}{\mathcal{V}}\sim e^{-B} \ , \quad B=\frac{8\pi^2 \Delta V}{3 H^4} \ , \label{eq:HMpathint}
	\end{equation}
	The HM solution describes the transition of a Hubble volume from the false vacuum to the top of the barrier. In a similar manner to the flat space case, this bounce solution describes a decay only when there is a single negative eigenvalue of $S''_{HM}$ which happens as long as
	\cite{Tanaka:1992zw}
	\begin{equation}
		H>\sqrt{|V''_\text{top}|}/2 \ . \label{eq:eigen}
	\end{equation}
	Thus the HM solution is expected to describe the decay rate at large curvatures, when the Coleman-de Luccia bounce doesn't exist or its properties are different from standard low curvature expectations. 
	
	\subsubsection{Thermal interpretation}
	When considering tunneling in a fixed de Sitter background, one can interpret the bounce solutions as thermally assisted tunneling.  This is possible since in de Sitter spacetimes we can define a temperature $T=H/(2\pi)$, due to the finiteness of its horizon. To compute the decay rate in dS, we will work within the thermal interpretation proposed by Brown and Weinberg \cite{Brown:2007sd} (based on previous results by \cite{Affleck:1980ac,Linde:1981zj,LINDE198137}). We start by considering the scalar field living in a fixed dS spacetime whose action can be thought as the thermal effective action where the thermal modes, in this case the gravitons, have been integrated out. This EFT description is appropriate to compute the tunneling rate since the bubble scale, $H$, is much smaller than the scale of the thermal modes, $M_{Pl}$. Intuitively, the tunneling process can be thought of as consisting of two parts: the first one corresponding to the thermal excitation of states localized in the false vacuum with $E>E_{\mathrm{fv}}$ and the second one the quantum tunneling. 
	
	Formally, the decay rate can be defined as a thermal average given by \cite{Affleck:1980ac}
	\begin{equation}
		\Gamma=\frac{1}{Z_\text{fv}} \int_0^\infty d E e^{-\beta E} \rho(E) \Gamma(E) \ ,
	\end{equation}
	where $\rho(E)$ is the density of states and $\Gamma(E)$ is the tunneling rate for a given energy $E$. At high temperatures (equivalently, high curvatures), the integral is dominated by the $E>V_\text{top}$ region, and the decay rate can estimated as \cite{Brown:2007sd}
	\begin{equation}
		\frac{\Gamma}{\mathcal{V}}^\text{high T.}\sim \int_{E_{\text{top}}}^\infty d E e^{-\beta E} e^{\beta E_{\mathrm{fv}}}\sim e^{-\beta(E_{top}- E_{\mathrm{fv}})} \ , \label{eq:BoltzSup}
	\end{equation}
	which agrees with the HM estimate computed from path integral methods in Eq.~\eqref{eq:HMpathint} when we identify the potential energy increment in a horizon volume as
	\begin{equation}
		\Delta E=E_{\mathrm{top}}- E_{\mathrm{fv}}=\frac{4 \pi}{3 H_{\mathrm{fv}}^3}\Delta V \ .
	\end{equation}
	From this, we see that the Hawking-Moss solution can be interpreted as a purely thermal transition where the thermal fluctuations push the field all the way to the top of the potential barrier and then rolls down to the true vacuum \cite{Weinberg:2006pc}. Thus we note that as we increase the temperature, we observe a transition from the Coleman-de Luccia to Hawking Moss driven decay rate.
	
	The result for the decay rate in Eq.\eqref{eq:HMpathint} is missing the prefactor which requires a more careful calculation. Nevertheless, we cannot use a straightforward generalization of Eq.~\eqref{eq:flatdecayrate} since the Hawking-Moss case does not have a clear analogy with flat space decays in the sense that there are no zero modes and the dilute gas approximation cannot be used. Instead, we will perform a more precise calculation by pushing forward the thermal interpretation. At high temperatures, the physics is driven by long-wavelength modes thus we can use a semi-classical approach in this regime. To perform this semi-classical approximation we use the flux-over-population method for computing escape rates \cite{Kramers:1940zz,LANGER1969258,Berera:2019uyp,Gould:2021ccf}. This calculation consists of solving the Fokker-Planck equation for the scalar field assuming an initial probability distribution localized in the false vacuum which evolves to the equilibrium state. The solution relies on the special boundary conditions given by the steady-state solution which assumes a source behind the false vacuum and a sink right after the top of the barrier so that there is a constant probability current across the barrier. This also guarantees that we only compute the decay of the false vacuum and we do not include contributions from the flux crossing the barrier back to the false vacuum. These extra contributions exist when there is a minimum on the other side of the barrier, that is, the true vacuum; but they don't appear if the potential is unbounded from below on the other side of the barrier. Note also that the temperature cannot grow indefinitely since we require that it is smaller than $\Delta E$ to have a well defined initial state localized at the false vacuum. The flux-over-population method leads to a decay rate that factorizes into an equilibrium and a non-equilibrium contribution \cite{Kramers:1940zz,LANGER1969258,Berera:2019uyp,Gould:2021ccf,Ekstedt:2022tqk}; thus, the one-loop HM decay rate is given by
	\begin{equation}
		\Gamma=
		\frac{\kappa}{2\pi}\left|\frac{\operatorname{det} S^{\prime \prime}\left(\phi_{HM}\right)}{\operatorname{det} S^{\prime \prime}\left(\phi_{\mathrm{fv}}\right)}\right|^{-1 / 2}  e^{-B} \quad \text{for} \quad 2\pi T > \kappa \ .
	\end{equation}
	The factor of $\kappa/(2\pi)$ is the non-equilibrium contribution, commonly referred to as the dynamical factor, where $\kappa$ is the growth rate of the unstable mode at the saddle-point. To compute the dynamical prefactor, we look at the scalar field equation of motion which is given by the following Langevin equation
	\begin{equation}
		\left(\partial_{t}^{2}-\nabla^{2}\right) \phi(\vec{x}, t)+\frac{\partial V(\phi)}{\partial \phi}+\eta \dot{\phi}(\vec{x}, t)=\xi(\vec{x}, t)\ ,
	\end{equation}
	where the damping coefficient is $\eta=3H$ and $\xi$ is the Gaussian white noise. Expanding the scalar field $\phi$ in a series around $\phi_{\text {top }}$ and taking the ansatz $\phi-\phi_{\text {top }}=C e^{\kappa t}$, one finds that
	\begin{equation}
		\kappa=-\frac{3}{2}H\left(1-\sqrt{1+\frac{4|V''_\text{top}|}{9H^2}}\right) \ . \label{eq:kappa}
	\end{equation}
	In obtaining the result above, we have assumed that the field gradients and the noise term are subdominant. In several applications where dissipation can be neglected, the dynamical factor can be shown to be given by the negative eigenvalue. This can not be done in our case because the dissipation, given by the damping coefficient $\eta$, is proportional to the Hubble parameter which cannot be much smaller than $\sqrt{|V''_\text{top}|}$, see Eq.~\eqref{eq:eigen}. Taking the small dissipation limit is equivalent to taking  $|V''_\text{top}|/ H^2\gg 1$. In this limit, $\kappa$ approaches the HM negative eigenvalue $\sqrt{|V''_\text{top}|}$, but the Hawking-Moss instanton does not describe tunneling anymore. One can also note that the HM-CdL transition occurs as we approach the limit $\kappa=H$. In the large temperature (large friction) limit we have $\ddot{\phi}\ll3H\dot{\phi}$ which leads to the simplified expression for the prefactor
	\begin{equation}
		\kappa^\text{high T}=\frac{|V''_\text{top}|}{3H} \ .
	\end{equation}
	
	We can further simplify the expression of the decay rate by writing the result explicitly
	\begin{equation}
		\Gamma=\frac{\kappa}{2\pi}\frac{\sqrt{V''_\text{fv}}}{\sqrt{|V''_\text{top}|}}\frac{\int \prod_{n\neq 0}\frac{\ud c_n^b}{\sqrt{2\pi}} e^{-S_b-\frac{1}{2}\sum_{n\neq 0}(c_n^b)^2 \lambda_n^b}}{\int \prod_{n\neq 0}\frac{\ud c_n^\textbf{fv}}{\sqrt{2\pi}} e^{-S_\textbf{fv}-\frac{1}{2}\sum_{n\neq 0}(c_n^\textbf{fv})^2 \lambda_n^\textbf{fv}}} \ ,
	\end{equation}
	where we have expanded the field as $\phi=\phi_b+\sum_n c_n\psi_n$, with $\psi_n$ are the eigenfunctions of $S''$, and rewritten the integration measure in terms of the coefficients $c_n$. Furthermore, we integrated over the homogeneous modes $\psi_0$. This expression can be rewritten in terms of the constraint effective potential \cite{FUKUDA1975354,ORAIFEARTAIGH1986653}, which is defined as
	\begin{align}
		e^{-\int\ud^4 x U^\text{1 loop}}&=\int \ud \phi \, e^{-S[\phi]} \,  \delta\left(\frac{1}{\mathcal{V}}\int\phi \ud^4 x - \phi_b\right)  \nonumber \\
		&=\int \prod_n\frac{\ud c_n}{\sqrt{2\pi}} e^{-S-\frac{1}{2}\sum_n c_n^2 \lambda_n}\delta(c_0) \ ,
	\end{align}
	where the delta function tells us to integrate over the inhomogeneous modes only. Note that because of the finite volume of the de Sitter space, different definitions of the effective potential are not equivalent.
	
	Thus, we can write the decay rate as
	\begin{align}
		\boxed{
			\Gamma=\frac{\kappa}{2\pi}\sqrt{\frac{V''_\text{fv}}{|V''_\text{top}|}} e^{-\frac{8\pi^2 \Delta U^\text{1 loop}}{3 H^4}}\ ,} \label{eq:DecayRateSC}
	\end{align}
	and in the high temperature limit ($|V''_\text{top}|/ (4 \pi^2)\ll T^2$), we find
	\begin{align}
		\Gamma^\text{high T}=\frac{\sqrt{V''_\text{fv}|V''_\text{top}|}}{2\pi} e^{-\frac{8\pi^2 \Delta U^\text{1 loop}}{3 H^4}}\ .  \label{eq:DRhighT}
	\end{align}
	The simplicity and validity of our analysis heavily relies on the fact that the Hawking-Moss solution is given by a constant saddle-point $\phi=\phi_\text{top}$. This allows for the simple interpretation of the escape rate as the tunneling rate, since both end points correspond to the top of the barrier. Similarly this avoids possible issues with double-counting of modes when using the constraint effective potential since our saddle-point is simply a constant \cite{Croon:2020cgk,Gould:2021ccf}.

	\begin{figure*}
	\begin{subfigure}[h]{0.47\textwidth}
		\includegraphics[width=\textwidth]{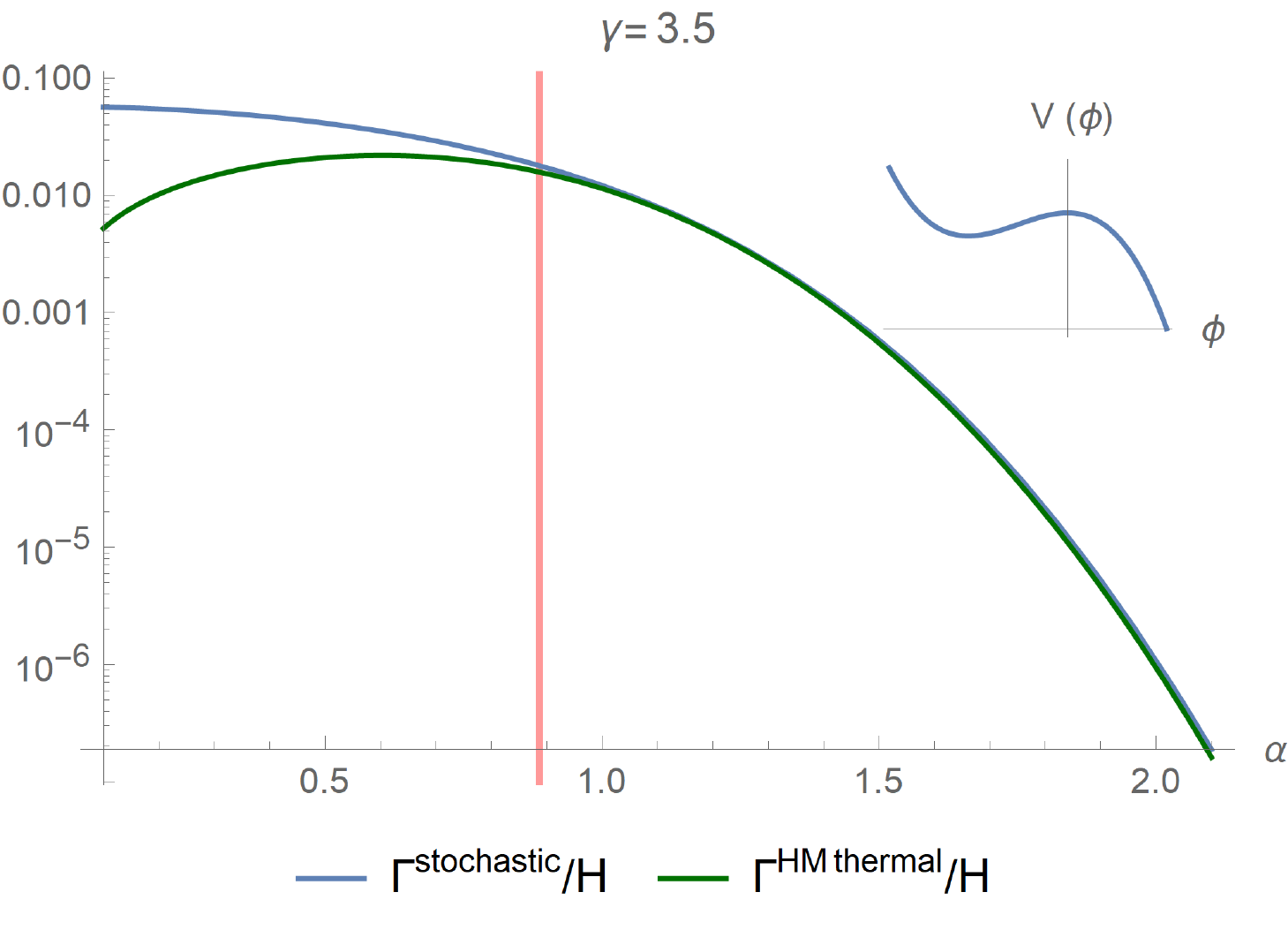}
		\caption{}
		\label{fig:phi3}
		\end{subfigure}
		~
    \begin{subfigure}[h]{0.47\textwidth}
    \includegraphics[width=\textwidth]{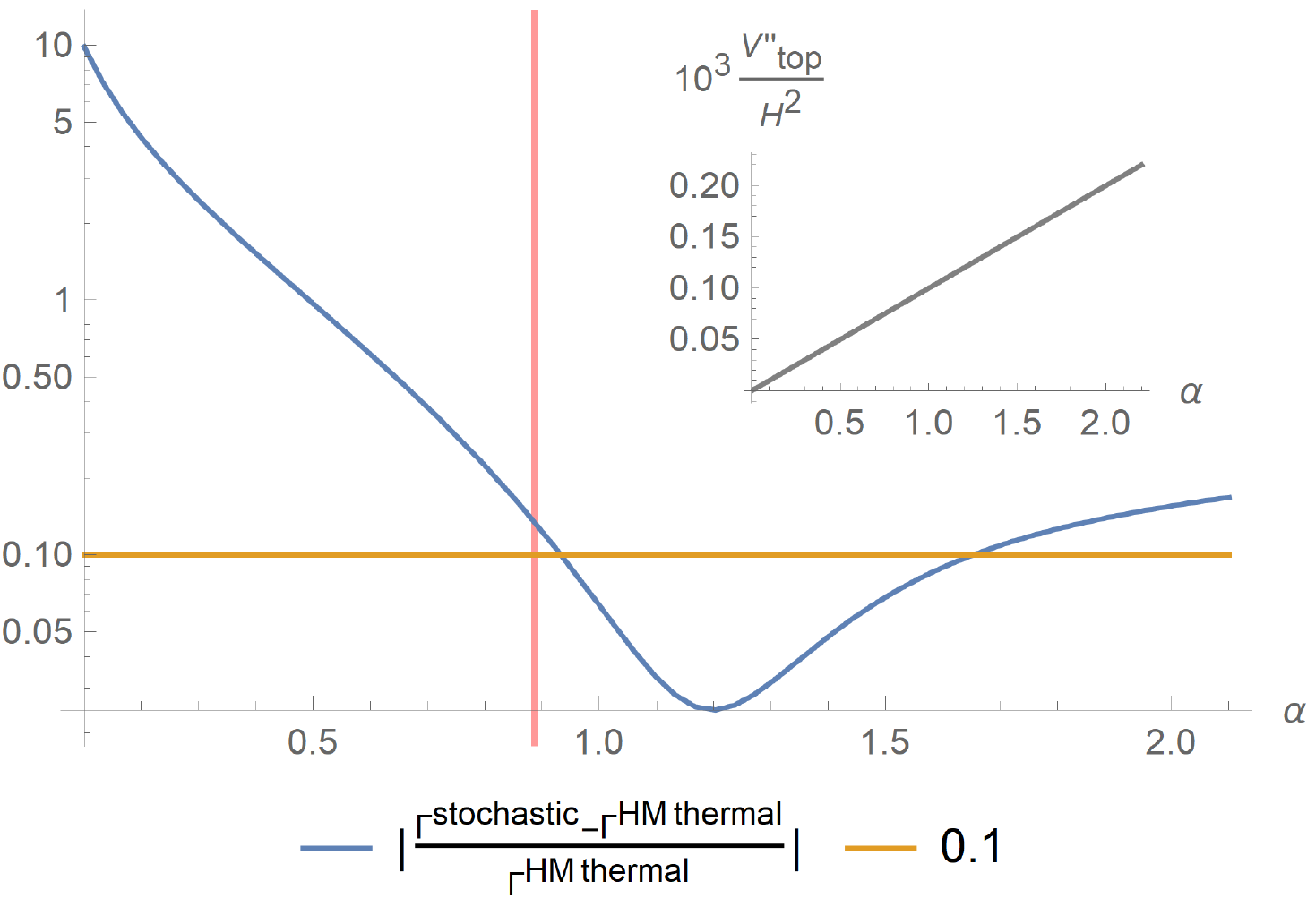}
	\caption{}
	\label{fig:errorphi3}
	\end{subfigure}
	\caption{Comparison of the decay rate for an unbounded potential with a metastable vacuum. We have taken the potential in Eq.~\eqref{eq:pot} with $\beta=\lambda=\Lambda_6=0$, $H=10^{-2} M_{Pl}$, and  $\lambda_5=10^{-4}$. For plot (a), the green line corresponds to Eq.~\eqref{eq:DecayRateSC} and the blue line is computed from the lowest eigenvalue of the Fokker-Planck equation in the stochastic approach. To the left of the vertical red line the saddle-point approximation breaks. In plot (b), we can observe that as $\alpha$ increases the discrepancy between the decay rates increases too. This is explained due to the increase in $V''_\text{top}$.}
    \end{figure*}
	
	One can note that Eq.~\eqref{eq:DecayRateSC} is the 0-dimensional escape rate for a scalar field in a fixed de Sitter spacetime with a potential given by the constraint 1-loop effective potential $U^\text{1 loop}$. This is a useful expression, since it helps us to compare the HM decay rate with the results from the stochastic formalism where the decay rate corresponds to the lowest eigenvalue of the FP operator. When there is only one direction in field space in which the decay can occur, the 0-dimensional escape rate corresponds to the stochastic formalism decay rate as long as we are within in the weak noise limit \cite{RevModPhys.62.251,PhysRevE.60.R1}. In the following section, we will show with explicit examples that the HM calculation agrees with the stochastic approach in the region where both results are applicable. One should note that the HM result is a saddle-point approximation. This approximation breaks down when perturbation theory ceases to be valid at either the top of the barrier or at the false vacuum. The latter corresponds to the region where $\Delta E<T$, that is, when the thermal fluctuations are large. In such region, we cannot have a metastable state localized in the false vacuum since the thermal fluctuations are large enough to destabilize it. Meanwhile the stochastic approach is not expected to be valid at large field masses due to the over-damped assumption. Outside of these regimes, both calculations of the decay rate should agree as long as we use the constraint effective potential in the stochastic approach. Namely, our analysis suggests that the stochastic approach can capture the one-loop corrections if we work with the constraint effective potential.

	\section{Comparison between Stochastic and Thermal approaches}

	In this section we will compare the results from the thermal HM decay rate calculation and the stochastic formalism one. For simplicity, we will consider that the one-loop constraint effective potential is given by
	\begin{align}
		U(\phi)=&3H^2M_{Pl}^2+\mu^{3} \phi-\frac{1}{2} m^{2} \phi^{2}-\frac{M}{3!}\phi^3 \nonumber \\
		&+\frac{\lambda}{4!} \phi^{4}-\frac{\lambda_5}{5!H}\phi^5+\frac{\lambda_6}{6!H^2}\phi^6 \ . \label{eq:pot}
	\end{align}
	In general, the shape of this potential will be given by a more complicated dependence on $\phi$ and should include a renormalization scale $\mu$ that arises after renormalizing the one-loop divergences. Nevertheless, this simple example is enough for our purposes of comparing the different calculations of the decay rate. We proceed to write the potential in terms of dimensionless variables. Defining
	\begin{align}
		&\alpha=\frac{m^2}{H^2 \lambda^{1/2}} \ , \ \ \beta=\frac{\mu^3}{H^3 \lambda^{1/4}} \ \ , \gamma=\frac{M}{H \lambda^{3/4}}  \ , \ \ \Lambda_5=\frac{\lambda_5}{\lambda^{5/4}}\ , \nonumber   \\ 
		&\Lambda_6=\frac{\lambda_6}{\lambda^{6/4}}\ , \ x=\Omega \lambda^{1/4} \frac{\phi}{H} \ , \quad \Omega=1+\sqrt{\alpha}+\beta+\gamma^{1/3} \ , \label{eq:dimless}
	\end{align}
	the rescaled potential reads
	\begin{align}
		\frac{v(z)}{\pi^2}=&\frac{4  M_{Pl}^2}{H^2}+\frac{4  \beta  z}{3 \Omega }-\frac{2  \alpha z^2}{3 \Omega ^2}-\frac{2  \gamma  z^3}{9 \Omega ^3} \nonumber \\
		&+\frac{ z^4}{18 \Omega ^4} -\frac{
   \Lambda_5 z^5}{90 \Omega ^5}-\frac{\Lambda_6 z^6}{540 \Omega ^6}\ . \label{eq:potential}
	\end{align}

\begin{figure*}
	\begin{subfigure}[h]{0.47\textwidth}
		\includegraphics[width=\textwidth]{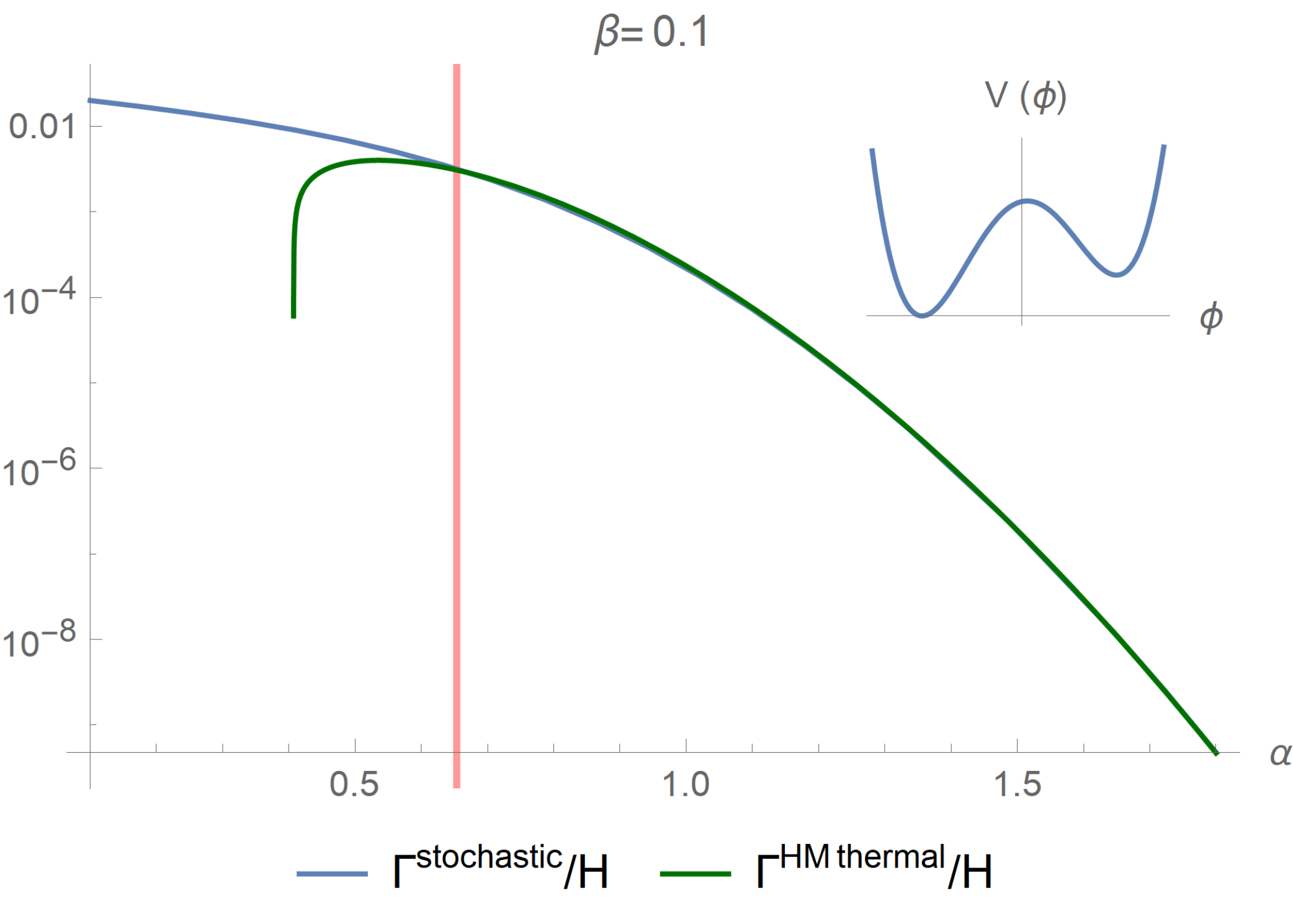}
		\caption{}
	\end{subfigure}
		~
		\vspace{0.4cm}
    \begin{subfigure}[h]{0.47\textwidth}
    \includegraphics[width=\textwidth]{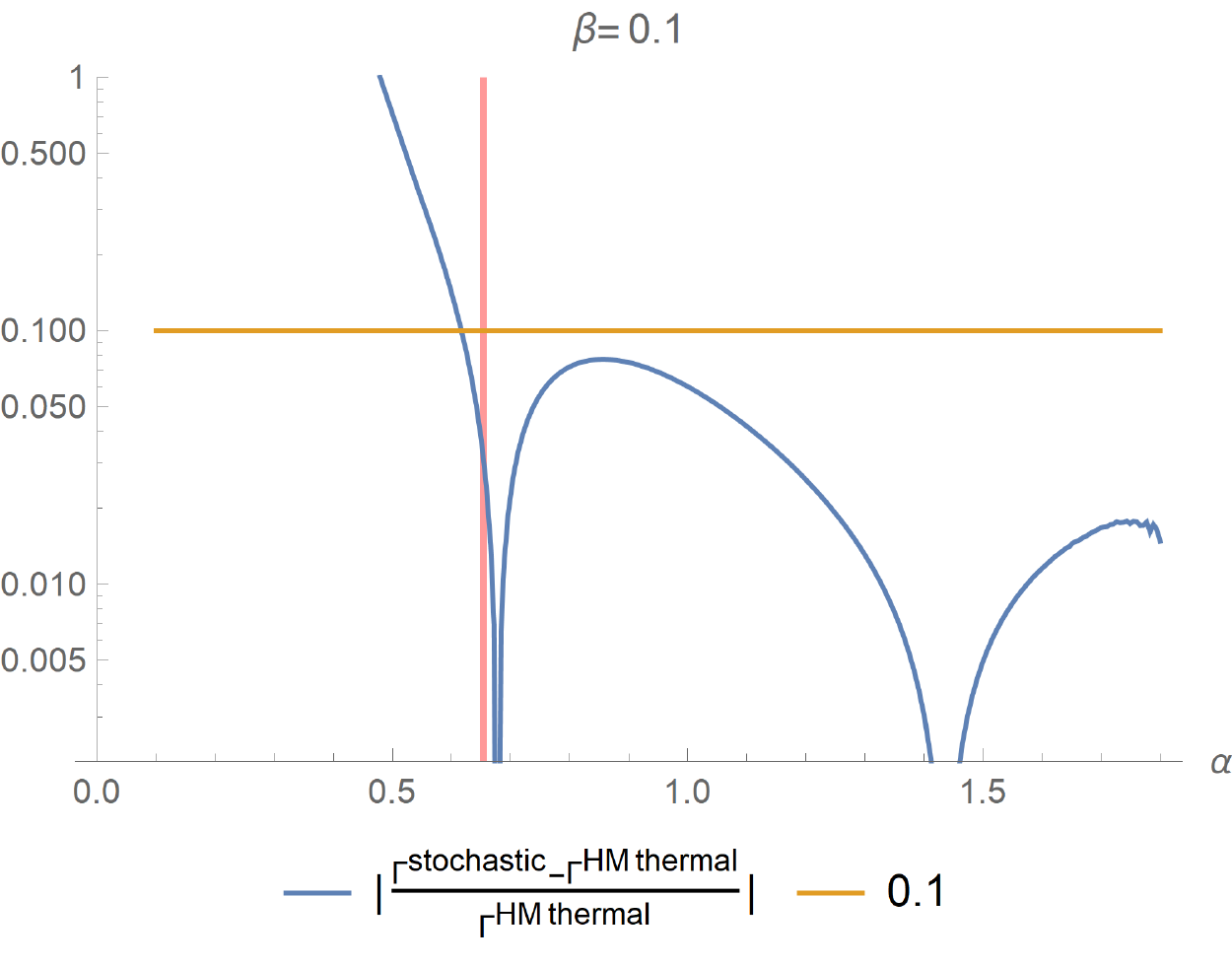}
    \caption{}
    \end{subfigure}
		~
		\vspace{0.4cm}
    \begin{subfigure}[h]{0.47\textwidth}
    \includegraphics[width=\textwidth]{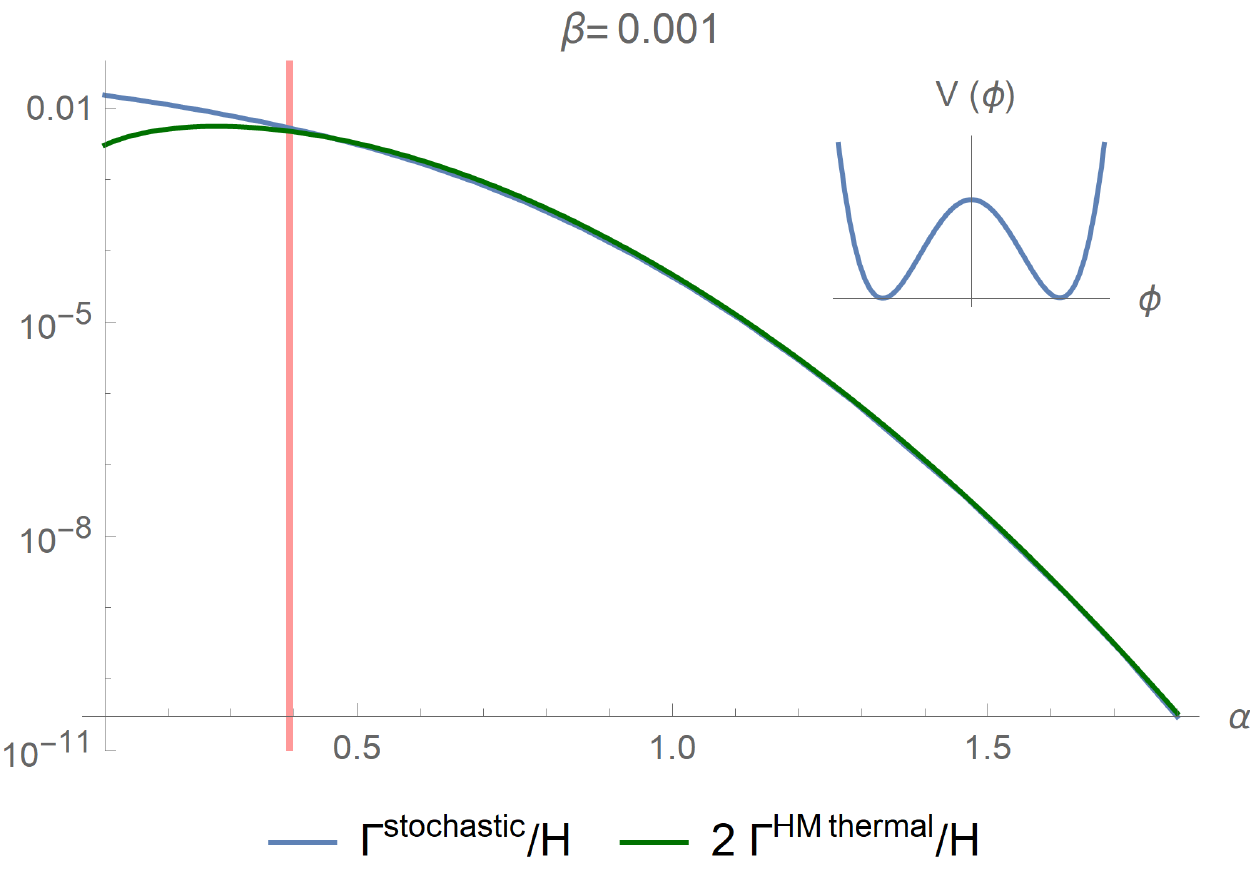}
    \caption{}
    \end{subfigure}
		~
    \begin{subfigure}[h]{0.47\textwidth}
    \includegraphics[width=\textwidth]{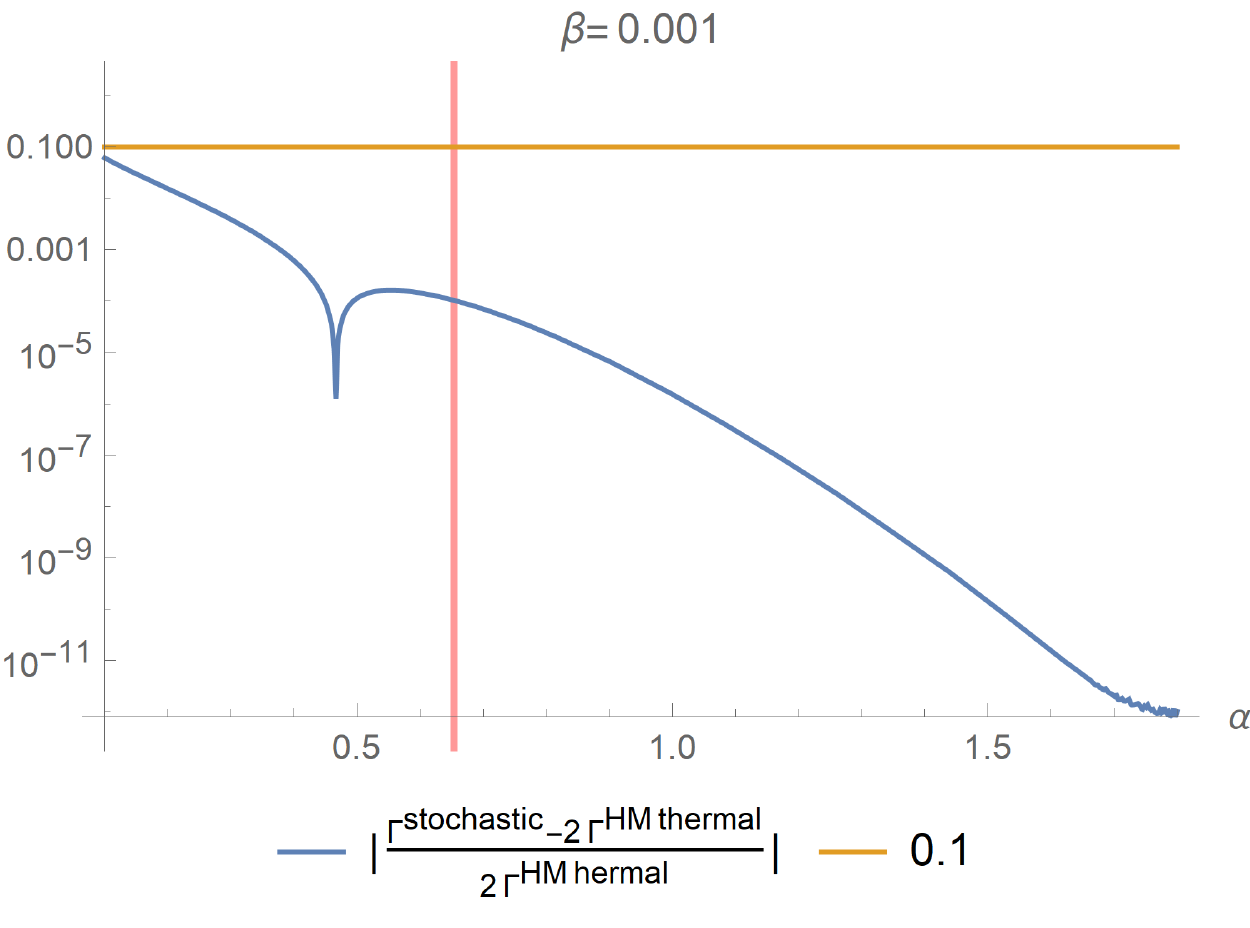}
    \caption{}
    \end{subfigure}
    \caption{Comparison of the decay rate of the false vacuum.  We have taken the potential in Eq.~\eqref{eq:pot} with $\gamma=\Lambda_5=\Lambda_6=0$, $H=10^{-3} M_{Pl}$, and $\lambda=0.05$. For plots (a) and (c), the green line is computed using Eq.~\eqref{eq:DecayRateSC} and the blue line is computed in the stochastic approach. Note that the sudden drop in the HM decay rate corresponds to the value of $\alpha$ where the potential ceases to have two vacua. To the left of the vertical red line (value at which $\Delta E =T$) the saddle-point approximation breaks. For plots (b) and (d), the blue line shows the discrepancy between the stochastic and instanton approach, we can see that for the present cases it is always small to the right of the red vertical line.
		}
		\label{fig:Comparisson}
	\end{figure*}

	We now analyze the decay rate for three different cases. The first one corresponds to a potential unbounded from below on one side, with a metastable vacuum where $\mu=\lambda=\lambda_6=0$. The dimensionless variables $\alpha$, $\gamma$, and $x$ are now defined as in Eq.\eqref{eq:dimless} with $\lambda=1$. The decay rate for this potential computed both from the thermal HM and stochastic approaches is shown in Fig.~\ref{fig:phi3}. We can see that the results agree for large masses which is the region where the thermal fluctuations are small and do not destabilize the unstable vacuum. In this region, the numerical calculation of the eigenvalues becomes difficult and and analytic expression such as Eq.~\eqref{eq:DecayRateSC} becomes useful. At small masses, when the potential barrier becomes increasingly small, we can no longer trust the thermal HM calculation. Meanwhile, the stochastic result still describes the decay rate as long as an unstable vacuum exists. One should note that at very large masses (large $\alpha$), the stochastic approach result becomes less precise, see Fig.~\ref{fig:errorphi3}. The massless limit assumed in the stochastic approach is equivalent to the high temperature limit result in Eq.~\eqref{eq:DRhighT}, but the precise value of the decay rate receives corrections that grow with the curvature of the potential at the top of the barrier, as seen in Eq.~\eqref{eq:kappa}.
	
	Since solving the FP equation is equivalent to solving the Schr\"odinger equation for a supersymmetric Hamiltonian \cite{Cooper:1994eh}, we can use well known results from zero-dimensional supersymmetry to understand the relation between the stochastic and the HM computations. In the stochastic approach, the decay rate is given by the lowest eigenvalue of the FP equation. Thus, given a potential $v$, the equation with the flipped potential, $-v$, will have the same lowest eigenvalue. This tells us that we can think of the decay rate as being computed for the flipped potential and that the decay rate should be symmetric under the exchange  $\text{false vacuum }\longleftrightarrow \text{top of the barrier}$, which we see that is indeed the case at large temperatures (Eq.\eqref{eq:DRhighT}).
	
	The second case is a potential with a true and a false vacuum where $\gamma=\lambda_5=\lambda_6=0$. We analyze this potential for different values of $\alpha$ and $\beta$ and show our results in Fig.~\ref{fig:Comparisson}. As in the previous case, the thermal HM calculation breaks down at small $\alpha$. Meanwhile, for large enough $\alpha$ and $\beta$ where there is a clear distinction between the false and true vacuum, we see perfect agreement between the HM and stochastic approach. When we decrease $\beta$, the vacua are nearly degenerate. The HM result still describes the tunneling from the false vacuum to the true vacuum, given the boundary conditions chosen to solve the FP equation when computing the escape rate. On the other hand, the stochastic approach includes the fluctuation bouncing back from true vacuum to the false vacuum. We can see this by looking again at the flipped potential $-v$. In that case there is a false vacuum (previously the top of the wall) with one wall on each side. Due to supersymmetry, the stochastic formalism will simply give the same result, while for the HM computation we have to account for the probability to transition to either side, or in other words double the decay rate. This is confirmed in the results from Fig.~\ref{fig:Comparisson} where we compare twice the value of the HM decay rate from Eq.~\eqref{eq:DecayRateSC}, with the numerical computation for the lowest eigenvalue of the FP equation.
	
\begin{figure}
		\centering
		\includegraphics[width=0.45\textwidth]{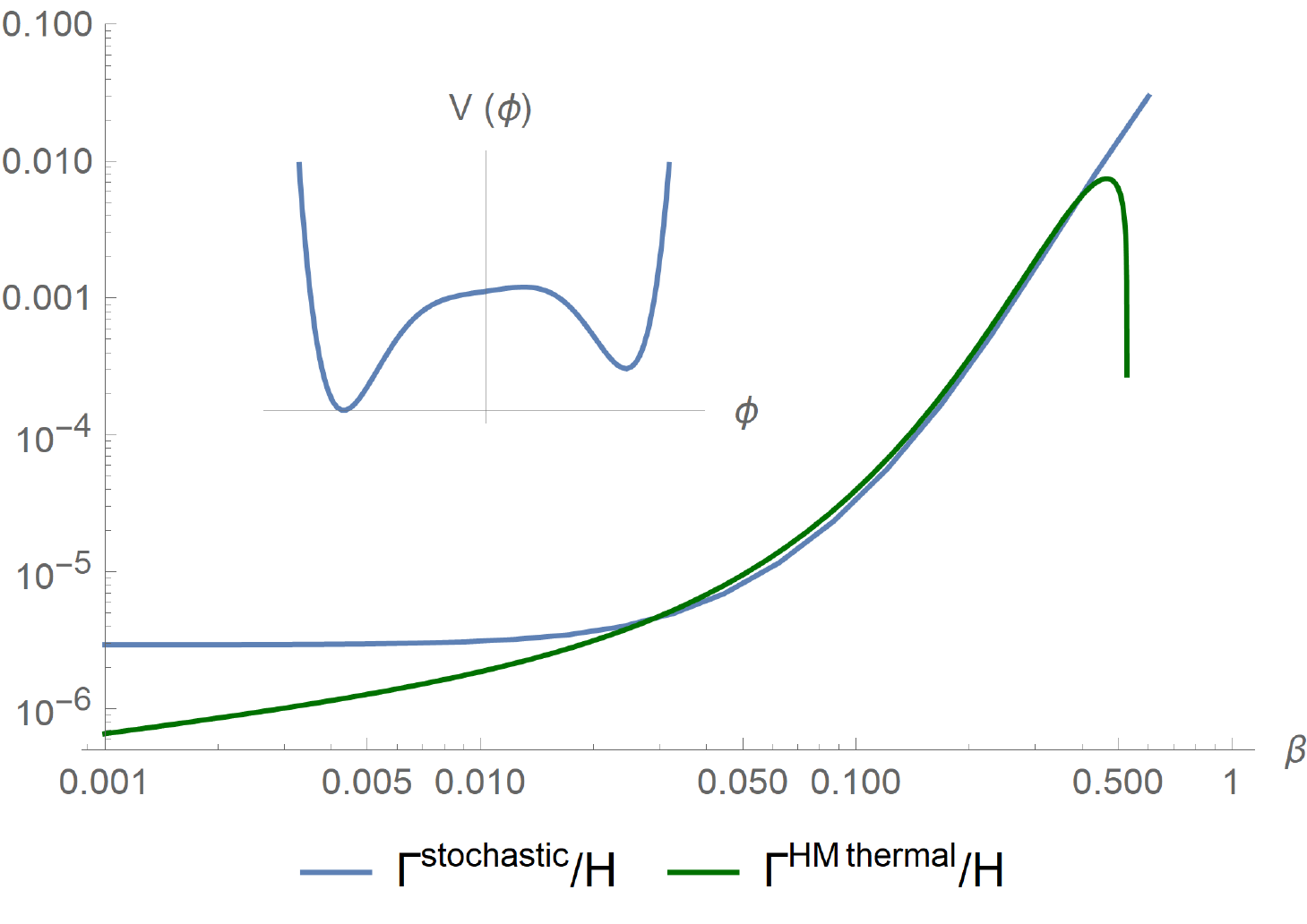}
		\caption{Breakdown of the thermal HM result. We have taken $H=10^{-3} M_{Pl}$, $\lambda=-0.3$, and $\lambda_6=60$. For small values of $\beta$, the path integral at the top of the barrier is non-gaussian due to the curvature becoming extremely small. At large $\beta$ the potential seizes to have two vacua; this happens when the HM result suddenly drops.}
		\label{fig:phi6}
	\end{figure}

	We consider a third case which will demonstrate the breaking of the HM result due to the non-gaussianity of the path integral at the top of the barrier. In this case, we take a potential with $\alpha=\gamma=\lambda_5=0$. The parameter $\beta$ measures the curvature at the top of the barrier, as we take $\beta\rightarrow 0$, $V''_\text{top}$ vanishes. Since we have $V''_\text{top}<<V^{(4)}_\text{top}$, perturbation theory, and hence the saddle-point approximation break. This is observed in Fig.\ref{fig:phi6} where the HM result largely deviates from the stochastic one as we decrease the coefficient of the linear term.
	
	Last, we analyze in detail the regions where the validity of the saddle-point approximation breaks. As a first step, we should find the region where $ \Delta E/T<1$. In principle, this requires knowledge of the tree-level potential, which we do not have for the present examples. Nevertheless, if we assume perturbation theory is valid then $ \Delta E/T \sim  \Delta E^\text{1 loop}/T$, where the 1-loop difference in energies is given by the difference between the constraint effective potential evaluated at the top of the barrier and at the false vacuum. Thus, the region where perturbation theory is valid and the saddle-point approximation holds corresponds to $ \Delta E^\text{1 loop}/T>1$. Another possibility for the  failure of the saddle-point approximation is the breaking of perturbation theory. For the first two examples analyzed here, we can see that the breaking of perturbation theory at the false vacuum happens when $T<E^\text{1 loop}\sim m^4/(\lambda H^3)$, that is at small $\alpha$. In fact, the lack of exponential suppression for the decay rate, which can be understood as thermal fluctuations becoming large and destabilizing the false vacuum, is equivalent to the non-gaussianity of the path integral at the false vacuum. In this regime, the decay rate computed from the HM bounce is not valid; which can be observe in Figs.~\ref{fig:phi3} and \ref{fig:Comparisson}. The second region where the saddle-point approximation breaks corresponds to the breaking of perturbation theory near the top of the barrier. In the third case analyzed above, this happens when the curvature at the top of the potential approaches zero (see Fig.~\ref{fig:phi6}). On the other hand, the stochastic approach assumptions break down at higher masses, i.e. higher $\alpha$, since the over-damped approximation is used when solving the Langevin equation. In this regime, the prefactor receives corrections that grow with the curvature at the top of the barrier as seen in Fig. \ref{fig:errorphi3}.
	
	\section{Discussion}
	In this article, we computed an explicit analytic formula for the decay rate in a de Sitter space at high Hubble rates using the Hawking-Moss instanton approximation including one-loop quantum corrections. We then showed that the stochastic Starobinsky-Yokoyama approach reproduces the Hawking-Moss result when the one-loop constraint effective potential is used instead of the classical potential, in the regime where both calculations are valid. It is important to note that because of the finite volume of the de Sitter space, different definitions of effective potential are not equivalent. For example, the more commonly used perturbative effective potential is not equal to the constraint effective potential. 
	
	Our results suggest that the stochastic approach with the constraint effective potential can give a non-perturbative way of computing vacuum decay rates when the saddle-point approximation, which the Hawking-Moss calculation relies on, is not valid. Correspondingly, we show that the stochastic method, which relies on the overdamped assumption, fails when the curvature of the potential becomes comparable to the Hubble rate. Meanwhile, the Hawking-Moss calculation is valid in this regime. Therefore, both methods are needed for a complete description of vacuum decay in de Sitter spacetimes at high Hubble rates.

	\section*{Acknowledgements}
	We would like to thank Oliver Gould and Stephen Stopyra for useful discussions. 
	All authors were supported by STFC grant ST/P000762/1, and MCG and AR also by the STFC grant ST/T000791/1. MCG was supported by the European Union's Horizon 2020 Research Council grant 724659 MassiveCosmo ERC--2016--COG. JEC was supported by the Carl Trygger Foundation through grant no. CTS 17:139.

	\bibliography{bib}
	
\end{document}